\def\etal{{\it et al.~\/}}

\def\ltsima{$\; \buildrel < \over \sim \;$}
\def\simlt{\lower.5ex\hbox{\ltsima}}
\def\gtsima{$\; \buildrel > \over \sim \;$}
\def\simgt{\lower.5ex\hbox{\gtsima}}

\documentstyle[12pt,aasms4,psfig]{article}
\tighten
\singlespace

\lefthead{A. Ferrara}
\righthead{}

\begin{document}
%\centerline{\tt PERSONAL COPY/ DRAFT FOR COMMENTS}
%\vskip 2 truecm

\title{The Positive Feedback of Pop~III Objects on Galaxy Formation}

\author{Andrea Ferrara$^1$}
\affil{
$^1$Osservatorio Astrofisico di Arcetri \\ 50125 Firenze, Italy 
\\ E--mail: ferrara@arcetri.astro.it} 
%.................................................................
\begin{abstract}
We study the formation of molecular hydrogen in cooling gas
behind shocks produced during the blow-away process thought
to occur in the first collapsed, luminous (Pop~III) objects  in the
early universe. We find that for a wide range of physical parameters
the $H_2$ fraction is $f \approx 6 \times 10^{-3}$. The  $H_2$ mass
produced in such explosions can exceed the amount of relic $H_2$ 
destroyed inside the photodissociation region surrounding a given
Pop~III. We conclude that these 
first objects, differently from the suggestion of Haiman \etal 1997,         
might have a net {\it positive} feedback on subsequent galactic 
formation. We discuss the effects of radiation and the implications
of our results for the soft-UV background.
\end{abstract}

\keywords{Cosmology: theory --  galaxy formation -- molecules
-- supernovae}

\section{Introduction}

Current models of cosmic structure formation based on CDM scenarios
predict that the first collapsed, luminous (hereafter Pop~III) objects should form at redshift $z\approx 30$ and have a total mass $M \approx 10^6 M_\odot$ or baryonic mass $ M_b \approx 10^5 M_\odot$
(Couchman \& Rees 1986, Haiman \etal 1997 [HRL], 
Tegmark \etal 1997). This conclusion is reached by requiring 
that the cooling time, $t_c$, of the gas is shorter than the Hubble time, $t_H$, 
at the formation epoch. In a plasma of primordial composition the only efficient 
coolant in the temperature range $T\le 10^4$~K, the typical virial temperature 
of Pop~III dark matter halos, is represented by H$_2$ molecules whose abundance increases 
from its initial post-recombination relic value to higher values during the halo 
collapse phase. It is therefore crucial to determine the cosmic evolution of such 
species in the early universe to clarify if small structures can continue to 
collapse according to the postulated hierarchical structure growth or if, 
lacking a cooling source, the mass build-up sequence comes to a temporary halt.

The appearance of Pop~III objects is now thought to cause a partial
destruction of the available molecular hydrogen either in the intergalactic medium (IGM) and/or in collapsing structures; the result is a negative feedback on galaxy formation. This effect has been pointed out by HRL   
and it works as follows. As stars form in the very first generation of objects, the emitted photons in the energy band 11.2-13.6 eV are able to penetrate the gas  and photodissociate
H$_2$ molecules both in the IGM and in the nearest collapsing structures, if they can propagate that far from their source. This negative feedback and its possible limitations are discussed by Ciardi, Abel \& Ferrara (1998, CAF). Here we propose and investigate instead a possible {\it positive} feedback based on supernova (SN) explosions, which, under many aspects, is reminiscent of a scaled version of the explosive galaxy formation scenario introduced by Ostriker \& Cowie (1981) and put forward by many others. Pop~IIIs are very fragile due to their low mass and 
shallow gravitational potential: only a few SNe are sufficient to blow-away (Ferrara 1998) their baryonic content and drive an expanding blastwave into the IGM, which eventually becomes radiative and allows the swept gas to cool in a dense shell. The cooling transient, as we will see, is characterized by a strong nonequilibrium condition in which recombination lags behind the temperature decrease. As already pointed out by 
Shapiro \& Kang (1987) and Kang \& Shapiro (1992), this is a favorable condition for H$_2$ formation. Our conclusions are that the amount of molecular hydrogen thus formed can
exceed the one destroyed via photodissociation, yielding a net increase of the H$_2$ 
in the universe. As a consequence, not only the galaxy formation process is not 
halted, but instead it is favored by the effects of Pop~III formation. 

Sec. 2 describes the main properties of multi-SN explosions in the early universe; Sec. 3 is devoted to the calculation of H$_2$ formation in their cooling shell. In Sec. 4  we compare the magnitude of positive and negative feedbacks; some discussion of the results is 
given in Sec. 5. 
 
\section{Properties of Pop~III Remnants}

The mechanical luminosity of an OB association in a Pop~III object
at redshift $z$  can be written as (Ciardi \& Ferrara 1997, CF) 
\begin{equation}
\label{L}
L={\epsilon_0\nu \Omega_b f_b \over\tau t_{ff}}M\simeq
4\times 10^{36}\Omega_{b,5} f_{b,8}(1+z)_{30}^{3/2} M_6 
{\rm ergs~s}^{-1};
\end{equation}
where $(1+z)_{30}=(1+z)/30$ and
$\epsilon_0=10^{51}$~ergs is the energy of a SN explosion;
we assume a Salpeter IMF, according to which one supernova 
is produced for each 56~$M_\odot=\nu^{-1}$ of stars formed.
The baryon density parameter is $\Omega_{b}=0.05 \Omega_{b,5}$, of
which a fraction $f_b \sim 0.08 f_{b,8}$ (Abel \etal 1997b) is able to cool 
and become available to form stars. The dark matter halo mass and
density are $M_6=M/10^6 M_\odot$ and $\rho \simeq 200 \rho_c = 200 
[1.88\times 10^{-29} h^2(1+z)^3]$~g~cm$^{-3}$, respectively (we assume a flat 
EdS universe $\Omega_m=1$ and $h=H/100$~ km~s$^{-1}$~Mpc$^{-1}$); 
the corresponding free-fall time is $t_{ff}= (4\pi G \rho)^{-1/2}$; 
$\tau^{-1}=0.6 \%$ is the star formation efficiency, calibrated on the 
Milky Way. The correlated SN explosions drive a blastwave in the 
surrounding gas, which eventually propagates into the IGM;
due to their low mass, the effects of Pop~III ISM on the blastwave
expansion can be neglected (MacLow \& Ferrara 1998). 
The evolution of the shock radius, $R_s$, can be obtained in the thin shell
approximation by solving numerically eqs. (5)-(6) of CF. For sake of 
simplicity we use the following (Sedov) analytical solution of the same equations which holds for the adiabatic case when the external confining pressure and the gravitational pull of the DM halo are neglected: 
\begin{equation}
\label{Rs}
R_s(t) \simeq 0.76 \left({E t^2\over \rho_b}\right)^{1/5},
\end{equation}
where $E=L t_{_{OB}}$, $t_{_{OB}}\approx 10^7$~yr is the average 
lifetime of massive stars, $E$ is the total energy of the explosion, 
$\rho_b=\Omega_b \rho$ the IGM density at redshift $z$.
Eq. \ref{Rs} does not take into account the effects of Hubble expansion,
which are negligible since $R_s \ll c/H$ (see eq. \ref{Rsc}). 

The corresponding temperature of the postshock gas is $27.6 (\dot 
R_s/{\rm km s}^{-1})^2$~K. Such hot gas will not be able to cool until 
its cooling time, $t_c$, becomes smaller than the Hubble time, $t_H$.  
The energy losses, $\Lambda(T)$, due to gas-radiative processes  in 
a primordial gas are of order $10^{-23}$~ergs~cm$^3$~s$^{-1}$ in the 
range $10^5 \le T \le 10^8$~K (Schmutzler \& Tscharnuter 1993); 
the other relevant cooling agent is 
inverse Compton cooling off CMB photons (Ikeuchi \& Ostriker 1986).  
Comparing the two rates, it is easily seen that inverse Compton dominates
the cooling for $z \simgt 185 \Omega_b h^2 -1$, for $T= 10^5$~K.
The condition $t_c \le t_H$ requires that  
$z \simgt 8h^{2/5} -1$; shocked gas produced at epochs earlier than this will
be able to cool and condense in a dense shell behind the shock front. 
Note that as the temperature drops to lower values with subsequent gas 
recombination, cooling via hydrogen Ly$\alpha$ line excitation 
will start to become important.
At $t=t_c= 1.2\times 10^3 (1+z)^{-4}$~Gyr, the shock radius and postshock
gas temperature are respectively
\begin{equation}
\label{Rsc}
R_s(t_c) = 0.16 f_{b,8}^{1/5}(1+z)_{30}^{-19/10} M_6^{1/5} h^{-1/5} 
{\rm kpc};
\end{equation}
\begin{equation}
\label{Tsc} 
T_s(t_c)= 4.9\times 10^4 f_{b,8}^{2/5}(1+z)_{30}^{21/5} M_6^{2/5} h^{-2/5}
{\rm K}.
\end{equation}
These relations hold if the Mach number of the
shock with respect to the IGM sound speed is ${\cal M}\gg 1$, a
condition verified if the IGM has not yet been reheated.

\section{Molecular Hydrogen Formation}

In the following, we calculate the time-dependent, 
nonequilibrium ionization and molecule abundance evolution 
behind the multi-SN shock as a function of the object 
mass and epoch of the explosion.
The postshock cooling gas is assumed to be 
initially at temperature $T_s(t_c)$ and density $4\rho_b$. 
In hydrogen of density $n = n_{H}+n_{H^+}+
n_{H_2}$  (we neglect He), the evolution of the ionization fraction,
$x= n_{H^+}/n$, molecular fraction, $f= n_{H_2}/n$, and
temperature, $T$, is determined by the following rate equations:
\begin{equation}     
\label{xrate}
\dot x = k_1 n x (1-x) - k_2 n x^2   
\end{equation}

\begin{equation}     
\label{frate}
\dot f = k_8 n_{H^-}(1-x) - k_{11} n x f - [k_{12}x + 
         k_{13}(1-x)] n f
\end{equation}

\begin{equation}     
\label{trate}
{\dot T} = -T\left[{(\gamma - 1)\over \gamma}{\Lambda(T)\over p} + 
{\dot x \over 1+x}\right].
\end{equation}
The various rate coefficients $k_i$ are labelled according to 
the nomenclature given in Abel \etal 1997a (Appendix A); 
$n_{H^-}$ is the $H^-$ species number density; 
$\gamma=5/3$ is the specific heat ratio and $p=(1+x)nk_BT$ is
the gas pressure.
In addition to the standard atomic cooling processes, 
$\Lambda(T)$ includes Compton and $H_2$ cooling;
the latter has been adopted from Martin \etal (1996). 
We assume that no heating is provided to the gas after it
has been heated by the shock to the initial temperature
and neglect adiabatic cooling due to Hubble expansion
for reasons similar to those given in the previous Section.

The first equation describes the usual collisional
ionization (rate $k_{1}$) -- recombination ($k_{2}$) balance. 
The equation for $f$ is slightly more complicated
and we describe it shortly. At the low densities considered here,
in the absence of dust, $H_2$ is formed in the gas phase mainly 
via the channel $H+ e^- \rightarrow H^- +h\nu$, at rate $k_8$ 
(formation via the $H_2^+$ channel, when
included, is found to be negligible in our case); it is then 
destroyed by one of the following mechanisms:  
(i) it is collisionally ionized by proton impacts ($k_{11}$); 
(ii) it is dissociated by collisions with electrons ($k_{12}$) 
and/or H-atoms ($k_{13}$). Abel \etal 1997a have shown that 
using the equilibrium value for $n_{H^-}$ is a very good 
approximation since the reactions determining the $H^-$ abundance
occur on a much shorter timescale than those responsible for
the $H_2$ chemistry; therefore we have calculated $n_{H^-}$
from their eq. (24). Also, we note that in the above form,
the rate equations correspond to the so-called minimal model
described in that paper. 
% Here we could discuss He, but it would be confusing for the
% reader and making only the paper more tedious..TSRBAP do not
% consider it either! 
The energy equation has been solved in the isobaric limit; 
this approximation is known to be appropriate for the evolution   
of Galactic supernova remnants (Cioffi \etal 1988), where 
most of the radiative phase is spent at constant pressure;
the same result was found to hold by Shapiro \& Kang (1987) and
Kang \& Shapiro (1992) for a wide range of shock velocities in
a cosmological context.
The numerical solution of the above rate equations are shown
in Fig. 1 for two different object masses ($M_6=1,10$) and 
initial redshift ($z=20,25,30$). Almost independently on the 
specific values of these parameters a high $H_2$ fraction ($f\approx 
6\times 10^{-3}$) is produced by redshift ten, with a
steep increase occurring when the gas has cooled down at $T \approx
20,000$~K. 
$H_2$ production is faster in high-$z$, high $T_s$ shocks, but the
peak abundance is practically the same in all cases; the final gas
temperature is in the range $300-500$~K. We have also experimented
a different $H_2$ cooling function (Hollenbach \& McKee 1979):
this produces a cooler final state ($T=80-200$~K) but only a marginally
different asymptotic value $f\approx 5\times 10^{-3}$.
For comparison, we have also plotted in Fig. 1 the Tegmark \etal 1997
``rule of thumb'' value $f = 5\times 10^{-4}$ at which $H_2$ cooling
becomes important. 

\section{Positive or Negative Feedback ?}

The results obtained in Sec. 3 show that typically at least a
$H_2$ fraction $f_6  = f/6\times 10^{-3}$ is formed after
explosive events leading to the blow-away of Pop~IIIs.
As discussed in the Introduction, HRL have argued that the same
objects could suppress the surrounding $H_2$ abundance due to 
their UV photodissociating radiation. In order to evaluate the
impact of Pop~IIIs on the subsequent galaxy formation, 
largely regulated by the availability of $H_2$ in this mass
(and redshift) range, it is useful to compare the $H_2$ 
mass production vs. destruction.
A lower limit to the amount of $H_2$ produced in an 
explosion is readily found to be equal to $M^+_{H_2}(z)\simeq  
(4\pi/3) \rho_b R_s(t_c)^3 (2f) $ or 
\begin{eqnarray}
\label{mh2+}
M^+_{H_2}(z)\simeq  
102 f_{b,8}^{3/5}\Omega_{b,5}(1+z)_{30}^{-27/10} M_6^{1/5} 
f_6 h^{7/10} {\rm M}_\odot. 
\end{eqnarray} 
As shown by CAF, the UV/ionizing radiation from Pop~III's 
massive stars previous to blow-away will produce both a HII region and a region 
of photodissociated intergalactic $H_2$ (radius $R_d$) in which 
the object is embedded. The radius $R_d$ can be defined by
requiring  that the photodissociation timescale ($t_d
\approx k_{27}^{-1}$, notation as in eqs. \ref{xrate}-\ref{frate})
is shorter than $t_H$. This condition yields the definition
$R_d = S_{LW}^{1/2} (1+z)^{-3/4} h^{-1/2}$, where $S_{LW}=
\beta S_i(0)$ is the UV photon flux in the $H_2$ Lyman-Werner (LW) bands
(11.2--13.6~eV), 
assumed here to be proportional to the flux of Lyc photons, $S_i(0)$, 
just before the massive stars explode.
The value of the constant $\beta$ depends somewhat on the IMF
and on the evolutionary stage of the stellar cluster, but its
value should be reasonably close to unity.
Paralleling eq. \ref{L}, we estimate $S_i(0)$ to be
\begin{equation}
\label{S}
S_i(0)={f_{uvpp} f_{esc} \Omega_b f_b \over\tau t_{_{OB}} m_p}M\simeq
10^{48} f_{uvpp,48} f_{esc,20} \Omega_{b,5} f_{b,8} M_6 
{\rm ~s}^{-1},
\end{equation}
where $f_{uvpp} = 48000 f_{uvpp,48}$ is the UV photon production
per collapsed proton efficiency (Tegmark \etal 1994) and $f_{esc}= 
0.2 f_{esc,20}$ is the escape fraction of such photons 
(absorption can be caused both by interstellar neutral H and/or 
dust). 
We have checked that this simple estimate is within 
a factor of 2 of the value obtained from the recently revised version 
of the Bruzual \& Charlot (1993) spectrophotometric code. 
The adopted reference value $f_{esc} = 0.2$ is an upper limit
derived from observational (Leitherer \etal 1995, Hurwitz \etal 1997) 
and theoretical (Dove \& Shull 1994) studies. It follows that
\begin{equation}
\label{rd}
R_d= 2.4 (\beta S_{48})^{1/2} (1+z)_{30}^{-3/4} h^{-1/2} 
{\rm ~kpc},
\end{equation}        
where $S_{48}= S_i(0)/10^{48} {\rm s}^{-1}$.
Thus, the ratio between the $H_2$ mass produced and destroyed 
by Pop~IIIs is
\begin{equation}
\label{m+m-}
{M^+_{H_2}(z)\over M^-_{H_2}(z)}= \left({f\over f_{IGM}}\right)
\left[{R_s(t_c)\over R_d}\right]^3.
\end{equation}
The post-recombination relic fraction of intergalactic $H_2$
is estimated to be $f_{IGM} \approx 2\times 10^{-6} h^{-1}$ 
(Palla \etal 1995; Anninos \& Norman 1996; Lepp, private communication).
The previous relation is graphically displayed in Fig. 2,
along with the values of $R_s(t_c)$ and $R_d$. From that plot
we see that objects of total mass $M_6=1$ produce more $H_2$
than they destroy for $z\simlt 25$; larger objects ($M_6=10$)
provide a similar positive feedback only for $z\simlt 15$,
since they are characterized by a higher $R_d/R_s(t_c)$ ratio.
However, since in a hierarchical model larger masses form 
later, even for these objects the overall effect should be
a net $H_2$ production.

\section{Discussion and Conclusions}
 
The results obtained in the previous Section could be modified 
by the fact that so far we have neglected the effects of the 
Pop~III stellar cluster radiation impinging on the cooled shell. 
In fact, even after all the
SNe have exploded blowing-away the gas, the coeval low mass stars will
continue to produce some residual flux in the energy range 
$0.755-13.6$~eV, relevant to the $H_2$ formation network. 
Two processes might have some effect in this context: (i) $H^-$
photoionization, and (ii) $H_2$ two-step photodissociation.

The first process occurs at a rate $k_{23} \approx 6\times 10^{-17} 
S_{LW}/R_s^2(t_c)$~s$^{-1}$ per $H^-$ atom, or $2.4\times 10^{-10}
\beta M_6^{3/5}$~s$^{-1}$ using eqs. \ref{Rsc}, \ref{S} and reference values for
the remaining parameters. The most effective $H^{-}$ destruction
process among the ones included is $H_2$ formation, whose rate in
the same case is $\sim 1.4 \times 10^{-9}$~s$^{-1}$. Thus, photoionization 
is negligible, and we have checked this conclusion numerically. 

We can calculate the effects of two-step photodissociation on the shell 
as follows. The shell $H_2$ column density is $N_{H_2}
\simeq  f M_s/m_p R_s(t_c)^2 =
1.5\times 10^{17} (1+z)_{30}^{11/10} M_6^{1/5}$~cm$^{-2}$ (we have
assumed that all the swept IGM mass is in the cool shell). Since the
critical column density for self-shielding is 
$N_{H_2}^{crit}=10^{14}$~cm$^{-2}$  (Draine \& Bertoldi 1997), the
shell is optically thick to dissociating radiation. Nevertheless, 
the flux in the LW bands, $J_d =S_{LW} h_{_P}/ 4\pi R_s^2$,
where $h_{_P}$ is the Planck constant,
will induce  a dissociation front in the shell;
$J_d$ is very likely dominating on a possible background radiation.
The propagation speed will be $v_d \simeq \xi \lambda_d/t_d$, where 
$\lambda_d= N_{H_2}^{crit}/n_{H_2}$ is the LW photon mean free path,
and $\xi \approx 3-5$ is a constant obtained by comparison with numerical 
results. Thus, the time required to dissociate the entire shell is 
$\simeq N_{H_2}/n_{H_2} v_d = 3\times 10^8 (J_{d,21} \xi)^{-1}N_{H_2,17}$~yr,
where $J_{d,21}=J_{d}/ 10^{-21} {\rm ergs~cm}^2{\rm s}^{-1}{\rm Hz}^{-1}
{\rm sr}^{-1}$, and $N_{H_2,17}= N_{H_2}/10^{17} {\rm cm}^{-2}$.
Since theoretical work (Bruzual \& Charlot 1993) suggests that $\beta$ drops to 
$\approx 3\times 10^{-3} \ll 1$ after all the massive
stars (the main contributors to the UV flux) have died, 
dissociating the entire shell will require a very long time, certainly 
larger than the Hubble time. 
%\begin{equation}
%\label{k27}
%k_{27} \simeq 10^8 J_d^{thick}= 10^8 {S_{LW} h_{_P}\over 4\pi R_s^2} \left(
%{N_{H_2}\over N_{H_2}^{crit}}\right)^{-3/4}, 
%\end{equation}
%where $h_{_P}$ is the Planck constant and $J_d^{thick}$ is the radiation
%flux intensity self-shielded by the shell $H_2$ column density $N_{H_2}$; the
%critical value $N_{H_2}^{crit}=10^{14}$~cm$^{-2}$ for the process has 
%been obtained by Draine \& Bertoldi (1997). The value of $N_{H_2}$ 
%is of order  $N_{H_2} \approx f M_s/m_p R_s(t_c)^2 = 
%1.5\times 10^{17} (1+z)_{30}^{11/10} M_6^{1/5}$~cm$^{-2}$ (we have
%assumed that all the swept IGM mass is in the cool shell). It follows
%that the photodissociation timescale ($\approx k_{27}^{-1}$) exceeds
%the Hubble time by a factor $\beta^{-1}$ at $z=30$. Thus, if 
%$\beta < 1$ as discussed above, this effect is of secondary importance,
%although not completely negligible. The uncertainties on the value of 
%$\beta$ could be 
%restricted in principle by an accurate stellar population synthesis study;
%however, our persisting ignorance on the IMF in the early universe
%would jeopardize such an effort.
Finally, Kang \& Shapiro (1992) pointed out that any external
{\it ionizing} radiation field (which could be provided by the hot postshock
gas and/or by the residual intermediate mass stars in the cluster)
tends to increase the final value of $f$, although introducing a
time delay due to a temperature plateau in the evolution characteristic
of photoionization heating. Thus, our estimate seems to provide a 
robust lower limit to the molecular hydrogen abundance.

%Similar enhancements of the $H_2$ fraction can also be triggered in 
%shocks occurring during merging events and/or gas cloud collisions
%in the early universe  (Struck-Marcell 1982). 
How can this $H_2$ enhanced gas be used for galaxy formation ?
First, regions in which Pop~IIIs are clustered enough for their shells 
to interact might become sites of active star/galaxy formation. 
Shell interactions appear to be necessary to make them gravitational 
unstable, since the temperature required for isolated fragmentation (Couchman
\& Rees 1986) is well below the values ($ \simgt 100$~K) found here.
Next, these intergalactic shells can be accreted by neighbor objects,
possibly increasing their $H_2$ abundance above the threshold  
$f\simgt 5\times 10^{-4}$ required for efficient cooling (Tegmark \etal 1997).
	
We conclude that Pop~III objects can produce
regions of considerably high molecular hydrogen abundance due to
multi-SN shocks propagating in the IGM. 
We have also seen
that the $H_2$ thus produced can exceed the amount of relic $H_2$ 
destroyed inside the photodissociation region surrounding a given
Pop~III. This occurrence suggests that these first objects         
might have a positive feedback on subsequent galactic formation.
Such conclusion is certainly
valid on a local scale, as defined by the radii of the two influence
spheres, $R_s$ and $R_d$, of a single object. However, our definition
of $R_d$ (see Sec. 4) does not exclude that some photodissociating
flux is present outside $R_d$. If many of these objects are present 
in the universe at the same epoch, they might contribute to  
an early soft-UV background which, illuminating an isolated, collapsing cloud,
could photodissociate and depress its $H_2$ content; this possibility has been discussed
by HRL. However, CAF have demonstrated that, once the {\it surviving relic IGM
fraction between photodissociated spheres} is taken into account, 
the universe opacity to LW           
photons before reionization becomes of order unity on scales larger 
than $R_d$, but smaller than the typical interdistance between Pop~IIIs 
at $z=20-30$. This implies that the soft-UV background is very weak due
to intergalactic absorption, and, as a consequence,
the proposed negative feedback considerably inhibited. 
Furthermore, most of the dissociating flux is absorbed by the thick $H_2$
shells discussed above.
Given that in general $R_d > R_i$,
where $R_i$ is the HII region radius of a Pop~III (CAF), clearing of 
intergalactic $H_2$ by photodissociated sphere overlapping will occur
before complete reionization. It is not clear at this stage if that
event occurs before the typical mass scale of the collapsing objects 
is such that cooling can proceed via Ly$\alpha$ only, thus considerably 
weakening the arguments in favor of a temporary halt of galaxy formation 
at high redshift.

%.......................................................................
\vskip 2truecm
We acknowledge  T. Abel, B. Ciardi, D. Cox and G. Field for discussions;
E. Corbelli, Z. Haiman, A. Loeb, F. Palla, O. Pantano, P. Pietrini, and 
Y. Shchekinov for comments;  thanks are due to L. Pozzetti for help with 
the spectrophotometric code.

\newpage
\begin{figure}[t]
%\centerline{\psfig{figure=h2plus_fig1.ps,height=8cm}}
\centerline{\psfig{figure=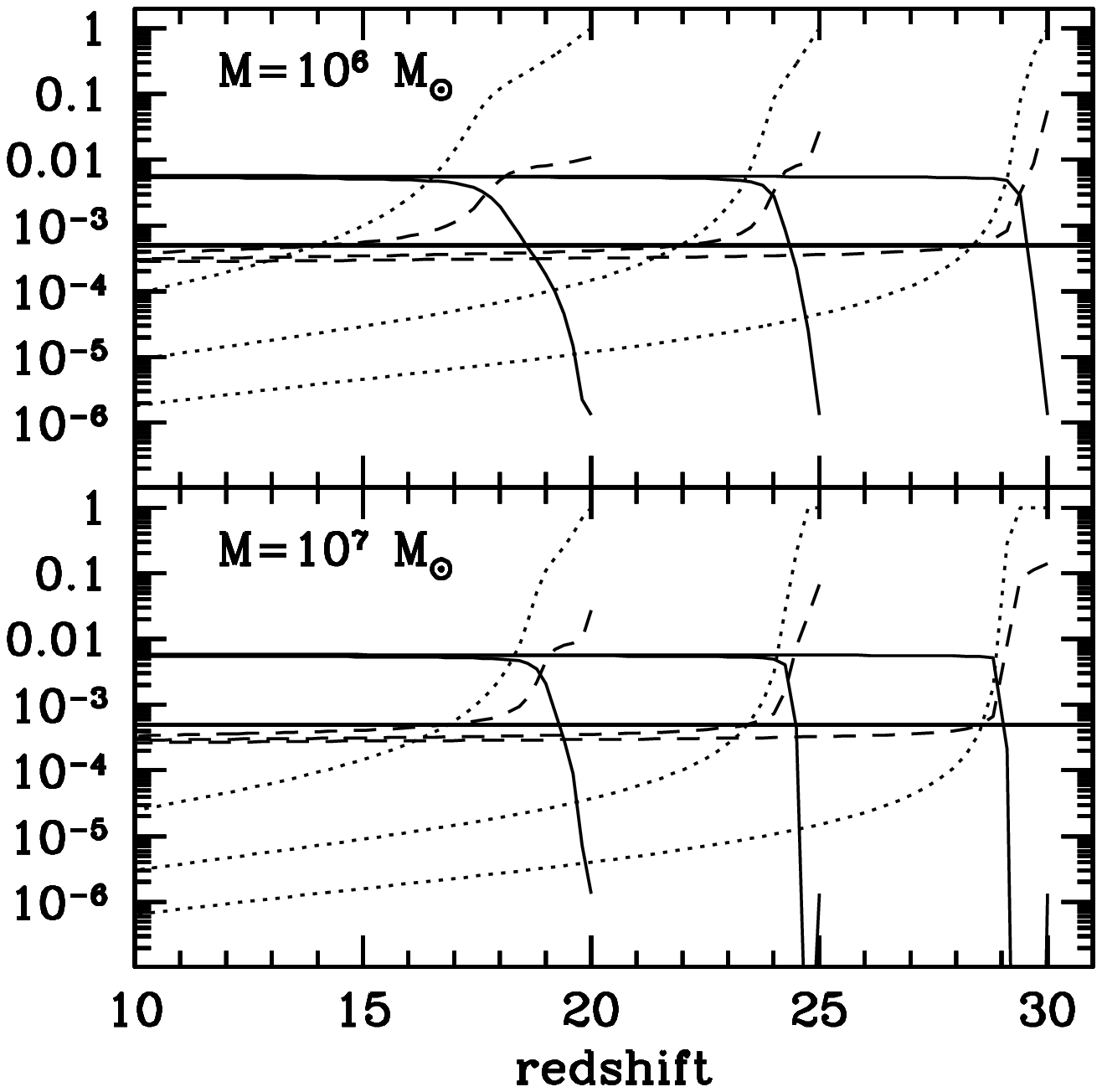}}
\caption{{Evolution of the $H_2$ molecular fraction $f$ (solid),
H ionization fraction $x$ (dotted) and temperature $T/10^6$~K 
(dashed) in the postshock gas for different values of the explosion 
redshift $z=20,25,30$ and Pop~III mass  $M=10^6 M_\odot$ (upper panel) 
and $10^7 M_\odot$ (bottom). The cosmological 
parameters are $\Omega_b=0.05$, $h=1$; the cooled baryon fraction is 
$f_b=0.08$. The thick horizontal line shows the Tegmark \etal 1997
threshold for efficient $H_2$ cooling.
}}
\end{figure}

\newpage
\begin{figure}[t]
%\centerline{\psfig{figure=h2plus_fig1.ps,height=8cm}}
\centerline{\psfig{figure=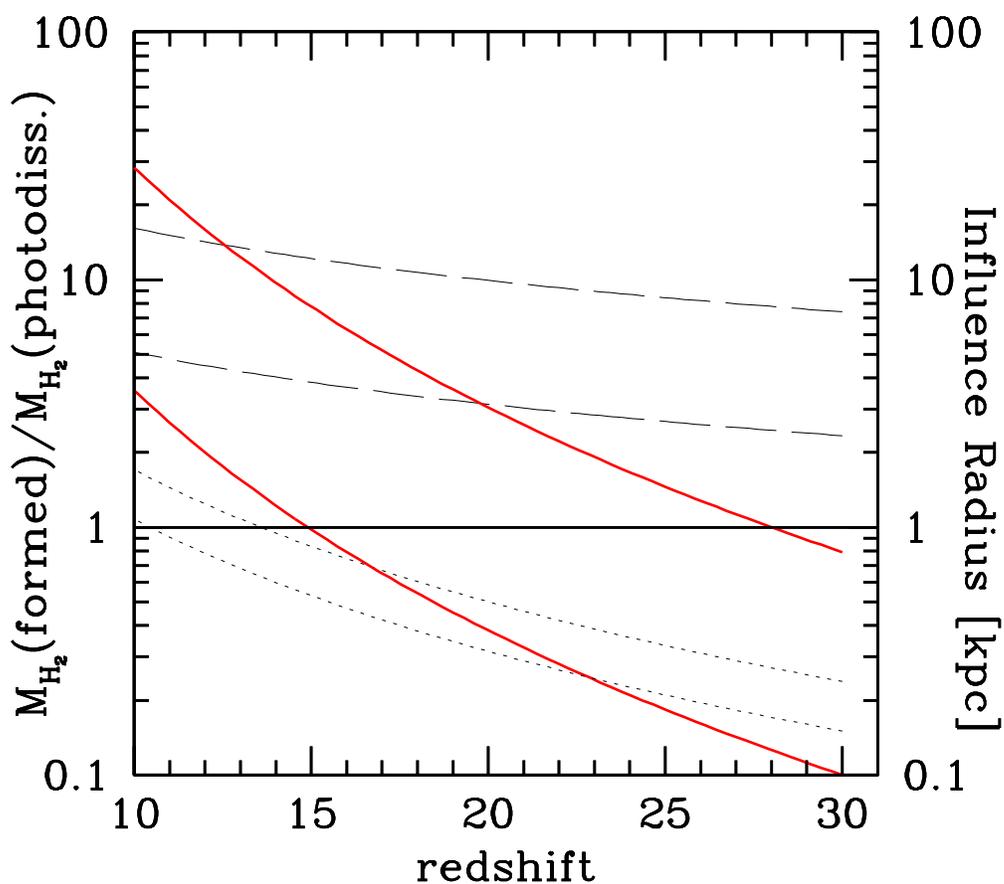}}
\caption{{Ratio between the $H_2$ mass formed
and destroyed by a Pop~III object as a function of the
multi-SN explosion redshift (solid lines); values larger
than one for the ratio define the epochs where Pop~IIIs
have a {\it positive feedback} on galaxy formation. Also shown are
the shell (proper) radius at cooling, $R_s(t_c)$ (dotted), and the 
photodissociation (proper) radius, $R_d$ (dashed). The upper set of curves
refers to objects of mass $M=10^6 M_\odot$, whereas the bottom one corresponds
to larger objects, $M=10^7 M_\odot$. The cosmological parameters are as in Fig. 1. 
}}
\end{figure}
\end{document}